\UseRawInputEncoding
\documentclass[letter,prapplied,
 amsmath,amssymb,
 reprint,
 author-year,%
superscriptaddress,
longbibliography
]{revtex4-1}

\usepackage{natbib}
\usepackage[export]{adjustbox}

\usepackage{float}
\usepackage{graphicx}
\usepackage{amsmath,amsthm,amssymb,amsfonts}
\usepackage{dcolumn}
\usepackage{bm}
\usepackage{gensymb}
\usepackage{physics}
\usepackage[normalem]{ulem}
\usepackage{hyperref}
\usepackage[caption=false]{subfig}
\graphicspath{ {images/} }
\usepackage[colorinlistoftodos]{todonotes}
\usepackage{wasysym}
\usepackage{array}
\usepackage{xcolor}
\usepackage[utf8]{inputenc} 
\usepackage{float}

\newcolumntype{C}{>{\centering\arraybackslash}X} 

\usepackage{cleveref}
\AtBeginDocument{%
    \newwrite\bibnotes
    \def\bibnotesext{Notes.bib}
    \immediate\openout\bibnotes=\jobname\bibnotesext
    \immediate\write\bibnotes{@CONTROL{REVTEX41Control}}
    \immediate\write\bibnotes{@CONTROL{%
    apsrev41Control,author="08",editor="1",pages="1",title="0",year="1"}}
     \if@filesw
     \immediate\write\@auxout{\string\citation{apsrev41Control}}%
    \fi
}%

\crefname{equation}{Eq.}{Eqs.}
\Crefname{equation}{Equation}{Equations}
\crefname{figure}{Fig.}{Figs.}
\Crefname{figure}{Figure}{Figures}
\crefname{section}{Sect.}{Sects.}
\Crefname{section}{Section}{Sections}
\crefname{table}{Table}{Tables}
\crefname{appsec}{Appendix}{Appendices}
\graphicspath{{Figures/}}
\usepackage{color, colortbl}
\definecolor{Gray}{gray}{1}
\definecolor{orange}{rgb}{1,0.97,0.9}
\definecolor{cyan}{rgb}{0.92,1,1}

\begin{document}

\title{Suppression of quasiparticle poisoning in transmon qubits by gap engineering}
\author{Plamen Kamenov}
\email{pok2@scarletmail.rutgers.edu}
\affiliation{Department of Physics and Astronomy, Rutgers University, Piscataway, NJ}
\author{Thomas J. DiNapoli}
\affiliation{Department of Physics and Astronomy, Rutgers University, Piscataway, NJ}
\author{Michael Gershenson}
\email{gersh@physics.rutgers.edu}
\affiliation{Department of Physics and Astronomy, Rutgers University, Piscataway, NJ}
\author{Srivatsan Chakram}
\email{schakram@physics.rutgers.edu}
\affiliation{Department of Physics and Astronomy, Rutgers University, Piscataway, NJ}
\date{\today}

\begin{abstract}
The performance of various superconducting devices operating at ultra-low temperatures is impaired by the presence of non-equilibrium quasiparticles. Inelastic quasiparticle (QP) tunneling across Josephson junctions in superconducting qubits results in decoherence and spurious excitations and, notably, can trigger correlated errors that severely impede quantum error correction. 
In this work, we use “gap engineering” to suppress the tunneling of low-energy quasiparticles in Al-based transmon qubits, a leading building block for superconducting quantum processors.  By implementing potential barriers for QP, we strongly suppress QP tunneling across the junction and preserve charge parity for over $10^3$  seconds. The suppression of QP tunneling also results in a reduction in the qubit energy relaxation rates. The demonstrated approach to gap engineering can be easily implemented in all Al-based circuits with Josephson junctions.

\end {abstract}

\pacs{Valid PACS appear here}
\maketitle

The reliable operation of many superconducting quantum devices relies on the assumption that fermion parity (i.e., the number of electrons modulo two) is conserved. This property is often referred to as the ``quantum purity" of superconductors at very low temperatures \cite{leggett1980a}. However, over three decades ago it was observed that the concentration of quasiparticles, $n_{QP}$, in thin-film superconducting devices at ultra-low temperatures could exceed that expected in thermal equilibrium by several orders of magnitude (see, e.g., \cite{joyez1994a,aumentado2004a,visser2011a,visser2014a}). The non-equilibrium quasiparticles (NQP) originate from Cooper-pair breaking by stray photons with energies above the superconducting gap $\Delta$~\cite{serniak2018a,rafferty2021spurious,gordon2022environmental,pan2022engineering, diamond2022}, high-energy phonons and electrons generated by cosmic rays~\cite{vepsaelaeinen2020a,mcewen2022resolving}, $\gamma$-ray sources in the qubit environment~\cite{cardani2021a}, and stress relaxation in substrates and films~\cite{mannila2022superconductor}. 

NQP adversely impact the performance of several types of superconducting quantum devices operating at milli-Kelvin temperatures, such as single-electron transistors \cite{kastner2000a}, quantum current standards based on Cooper pair pumps \cite{pekola2000a}, and microwave kinetic inductance detectors \cite{day2003a}. QP ``poisoning" is also expected to severely limit the coherence of devices designed for the realization of Majorana qubits~\cite{rainis2012a} and Andreev qubits~\cite{janvier2015coherent, hays2021coherent}. Most notably, the presence of NQP limits the performance of superconducting quantum bits based on Josephson junctions (JJ) \cite{martinis2021a}. Elastic QP tunneling results in dephasing of charge-sensitive qubits, while inelastic QP tunneling might be a significant source of qubit energy relaxation and excitation \cite{catelani2011a, Glazman2021}. Correlated errors associated with NQP “bursts” due to high-energy events make the task of quantum error correction - which relies on qubit errors being independent and individually below a certain threshold - very challenging \cite{wilen2021a,mcewen2022resolving}.
 
To overcome QP poisoning, two main approaches have been developed: (a) reducing the NQP density by shielding devices from stray photons and high-energy particles, or downconversion of QP-generating phonons~\cite{Iaia2022-hf} and (b) keeping the NQP away from the JJs by creating potential barriers that the NQP cannot overcome. The latter approach is realized by the special modulation of the superconducting energy gap in the junction electrodes, referred to as “gap engineering”~\cite{aumentado2004a,sun2012b,riwar2016a}. 

Gap engineering can be achieved by a number of methods, including introducing additional lower or higher-gap materials~\cite{wang2014measurement}, disorder~\cite{zhang2019microresonators}, or modulation of the film thickness~\cite{aumentado2004a}. This technique is expected to be efficient only when the barrier height significantly exceeds typical QP energies. Fortunately, the NQP created by high-energy events quickly relax toward the superconducting gap edge $\Delta$ due to inelastic electron-electron and electron-phonon collisions~\cite{gershenson2001a}. Near the gap edge, energy relaxation slows down considerably~\cite{lenander2011a,gershenson2001a,mannila2022a}, and the QP are long-lived due to slow recombination at low densities.

In this Letter, we report the successful implementation of gap engineering to suppress QP poisoning in Al-based transmon qubits~\cite{koch2007a,paik2011a}. We demonstrate that by creating potential barriers for NQP through the modulation of $\Delta$ in aluminum electrodes of Josephson junctions, we can suppress QP tunneling and maintain charge parity in transmon qubits for time intervals exceeding $10^3$ seconds. We realize this gap modulation by decreasing the film thickness of Al to increase $\Delta$ ~\cite{chubov1968a,meservey1972a,yamamoto2006a}, and demonstrate strong suppression of QP tunneling even with a relatively small modulation of $\delta\Delta\sim0.5\,\mathrm{K}$, where we set $k_B=1$. Notably, gap engineering enhances the coherence of our transmon qubits: the energy relaxation time $T_1$ markedly improves with the suppression of QP poisoning in our devices.

\begin{figure}
    \centering
    \includegraphics[width=\linewidth]{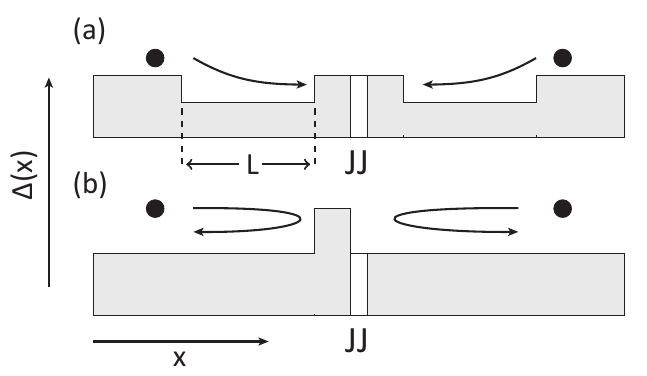}
    \caption{(a) The gap profile of a junction protected by two quasiparticle ``traps". The required size of the trap $L$ becomes prohibitively long in the devices with gap modulation in Al films due to the macroscopic ($\sim 100\,\mathrm{\mu m}$) energy relaxation length of low-energy quasiparticles. (b) The gap profile of a junction protected by a potential barrier. The required size of the high-gap barrier can be as thin as $(3-5) \xi$, where $\xi$ is the coherence length in a high-$\Delta$ film.}
    \label{fig:QP poisoning}
\end{figure}
There are two categories of engineered superconducting gap profiles that can be used for the suppression of QP poisoning, as illustrated in Figure \ref{fig:QP poisoning}: (a) QP "traps" \cite{riwar2016a} and (b) a QP potential barrier. In QP traps, NQPs relax to the gap edge in the region with low $\Delta$ compared to the rest of the device, potentially leading to QPs having insufficient energy to tunnel across the junction sandwiched by regions of higher $\Delta$. For the trap to work, the extent of the low-$\Delta$ region must significantly exceed the energy relaxation length for QPs with energies $\delta\Delta$, where $\delta\Delta$ is the gap difference between the high- and low-$\Delta$ regions. At low QP energies, the QP diffusion length, $L_\epsilon =\sqrt {D\tau_\epsilon}$, becomes macroscopic \cite{martinis2009a}, requiring sizable QP trap dimensions --- up to $L_\epsilon\sim300\,\mathrm{\mu m}$ for a QP energy $0.5\,\mathrm{K}$ above the gap edge, given a typical diffusion constant $D=0.01\,\mathrm{m^2/s}$ and an energy relaxation time $\tau_\epsilon \sim 10\,\mathrm{\mu s}$ \cite{gershenson2001a}. 

In contrast, in the case of a QP barrier, a small region of a material with high $\Delta$ is used, as shown in Fig. \ref{fig:QP poisoning}b. Here, the barrier can be as thin as $(3-5) \xi$, where $\xi$ is the coherence length in a high-$\Delta$ film. Considering that a typical value for $\xi$ in thin Al films does not exceed $\sim 0.1\,\mathrm{\mu m}$, the barrier length can be as small as $1\,\mathrm{\mu m}$.
Moreover, while it is essential for QP traps to be positioned on both sides of a JJ, it is sufficient to have a QP barrier only on one side of the junction. Therefore, in Al-based devices, barriers compare favorably with traps for the mitigation of QP poisoning. It's worth noting that if the gap difference $\delta\Delta$ is large (e.g., $14~\mathrm{K}$ for the pair Nb-Al), the length of QP traps may be as small as that for the barriers due to a steep dependence of $\tau_\epsilon$ on QP energies ($\mathrm{\tau_\epsilon}~(14 \mathrm{K}) \sim 10~\mathrm{ps}$ \cite{gershenzon1983quantum}).

While gap engineering has been successfully applied to suppress QP poisoning in a variety of devices based on Cooper-pair boxes (CPB) \cite{yamamoto2006a, bell2012a,joyez1994a,kalashnikov2020a,catelani2022a}, its application in the more ubiquitous transmon qubit encounters a number of challenges. For CPB devices, which typically feature two Josephson junctions in series separated by a small island, the required profile for gap engineering can be created straightforwardly by using a thin (20-25 nm) Al film for the CPB island, complemented by a thicker Al film for the remainder of the circuit. Additionally, these devices have a sizable charging energy, typically between $2-20\,\mathrm{GHz}$, which effectively raises the barrier for QP tunneling onto a CPB island. In contrast, the transmon qubit has a significantly lower charging energy, typically below $200\,\mathrm{MHz}$, and contains either single Josephson junctions or junctions configured in a SQUID geometry. Therefore, the fabrication process requires careful adjustments to ensure the successful formation of a QP barrier in a transmon.

\begin{figure}
    \centering
    \includegraphics[width=1\linewidth]{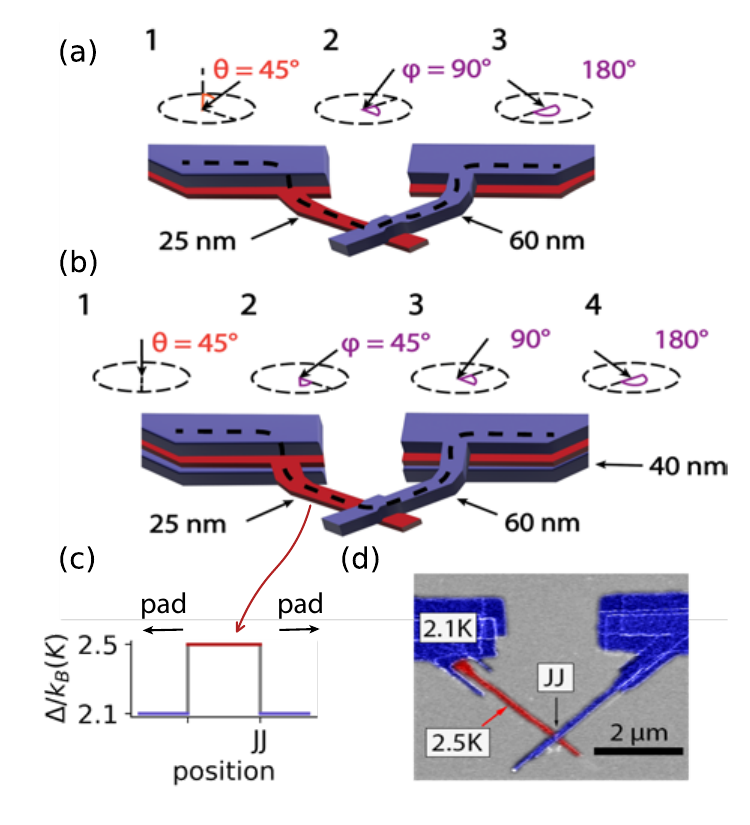}
\caption{Schematic diagram for the layers of the fabricated devices and their evaporation angles: (a)  for the unprotected ($\mathcal{NP}$) device without an underlayer. All deposition directions are at an angle of $\theta=45^\circ$ relative to the substrate. The bottom electrode ($t = 25\,\mathrm{nm}$) is deposited along direction 1, and the top electrode ($t = 60\,\mathrm{nm}$) is deposited in two steps ($30 + 30\,\mathrm{nm}$) along two opposite directions 2,3 for better step coverage. $\varphi$ indicates the azimuthal substrate rotation angle. (b) Schematic diagram for protected ($\mathcal{P}$) device with an additional underlayer. The underlayer (thickness $t = 40\,\mathrm{nm}$) is deposited along direction 1, followed by the three steps in (a). (c) The gap profile along the dashed line in panel (b). The Josephson junction is positioned at $l=3\,\mathrm{\mu m}$. The length of a larger-$\Delta$ region, $\sim3\,\mathrm{\mu m}$, is much greater than the coherence length in thin Al films, $\xi\sim0.1\,\mathrm{\mu m}$. (d) A microphotograph of the junction area with false colors corresponding to the value of the gap in Kelvin.}
    \label{fig:design}

\end{figure}

We implement gap engineering by utilizing the thickness dependence of the critical temperature in Al films. Pristine aluminum films below $\sim 30 \,\mathrm{nm}$ show a steep increase in $T_c$ \cite{chubov1968a,meservey1972a,yamamoto2006a}. The devices were fabricated on intrinsic Si substrates using a three-step electron-beam deposition of aluminum through a lift-off ``Manhattan pattern" as outlined in Fig.~2. Crucially, we introduce an additional low-gap Al underlayer in the first evaporation step and engineer a thinner $3\,\mathrm{\mu m}$ stretch with a higher gap in the bottom electrode. The Josephson junction was formed between the thin portion of the bottom electrode and the top electrode. The gap profile along the dashed line in Fig.~2b is shown in Fig.~2c, revealing a potential barrier on one side of the JJ which prevents NQP tunneling from the top electrode and keeps the NQPs generated in the wide portion of the bottom electrode away from the junction.  For additional fabrication details, see Section A of the Supplementary Information and Ref.~\cite{gladchenko2009a,bell2012a}).

To probe QP tunneling across the JJ, we fabricated transmon devices with relatively low ratio $E_J/E_C =15-20$, where $E_J$ is the Josephson energy and $E_C$ is the charging energy. The frequency $f_{ge}$ of
the transition between the ground ($\ket{g}$) and the first excited
($\ket{e}$) states of these charge-sensitive transmons depends on the charge parity. The difference in the values of  $f_{ge}$ for different charge parities in the charge-sensitive qubits was typically $10 - 30\,\mathrm{MHz}$, allowing the detection of QP poisoning events \cite{sun2012a,serniak2018a}. Several control devices were also fabricated (see Table 1): (a) devices without an underlayer (where there was no gap modulation in the bottom electrode), and (b) charge-insensitive devices operating in a conventional transmon regime with larger values $E_J/E_C$. Below we refer to the devices with and without an underlayer as the QP-protected ($\mathcal{P}$) and non-QP-protected ($\mathcal{NP}$), respectively.

\begin{table*}[t!]

\centering
\caption{Sample parameters, with sample names indicated by sample number and $\mathcal{P}$ for gap-engineered and $\mathcal{NP}$ for non-gap-engineered devices.  $f_{ge}$ is the range of frequencies of the transmon $ge$ transition over the range of offset charge ($n_g$) and $\epsilon_{ge}$ is the measured charge dispersion. The charging energy $E_C$ and the Josephson energy $E_J$ were determined by matching the $ge$ and $ef$ transition frequencies and charge dispersions to simulations conducted using scqubits~\cite{groszkowski2021scqubits}, with the coupling strength $g$ to the readout cavity simulated using black-box quantization~\cite{nigg2012black} and the energy-participation method~\cite{minev2021energy} and corroborated with the measured dispersive shift. The $T_1$ and $T_2$ values are quoted at base temperature.}

\begin{tabular}{|l|ccccccc|}
    \hline  
    Sample 
    & $E_J (\mathrm{GHz})$ & $E_C (\mathrm{GHz})$ & $E_J/E_C$& $f_{ge} (\mathrm{GHz})$ & $\epsilon_{ge} (\mathrm{GHz})$ & $T_1 (\mathrm{\mu s})$ & $T_2 (\mathrm{\mu s})$   \\
    \hline \hline
    1 $\mathcal{NP}$ 
    & 21.67 & 0.150 & 144 & 4.95 & --          & 12    & 	6.2      \\
    2 $\mathcal{NP}$
    & 7.417 & 0.403 & 18.4 & 4.438-4.448 & 0.010  & 18    &	1.2 \\
    \rowcolor{cyan}
    1 $\mathcal{P}$  
    & 13.69 & 0.150 & 91.3 & 3.897 & --          & 35-55 &	6-10       \\
    \rowcolor{cyan}
    2 $\mathcal{P}$  
    & 6.92 & 0.429 & 16.1 & 4.380-4.402 & 0.022  & 25-35 &	4.5  \\
    \rowcolor{cyan}
    3 $\mathcal{P}$ 
    & 5.92 & 0.400 & 14.8 & 3.887-3.913 & 0.026  & 40    &    1.0    \\
    \hline

\end{tabular}
\end{table*}

\begin{figure}[t!]
    \centering
    \includegraphics[width=0.5\textwidth]{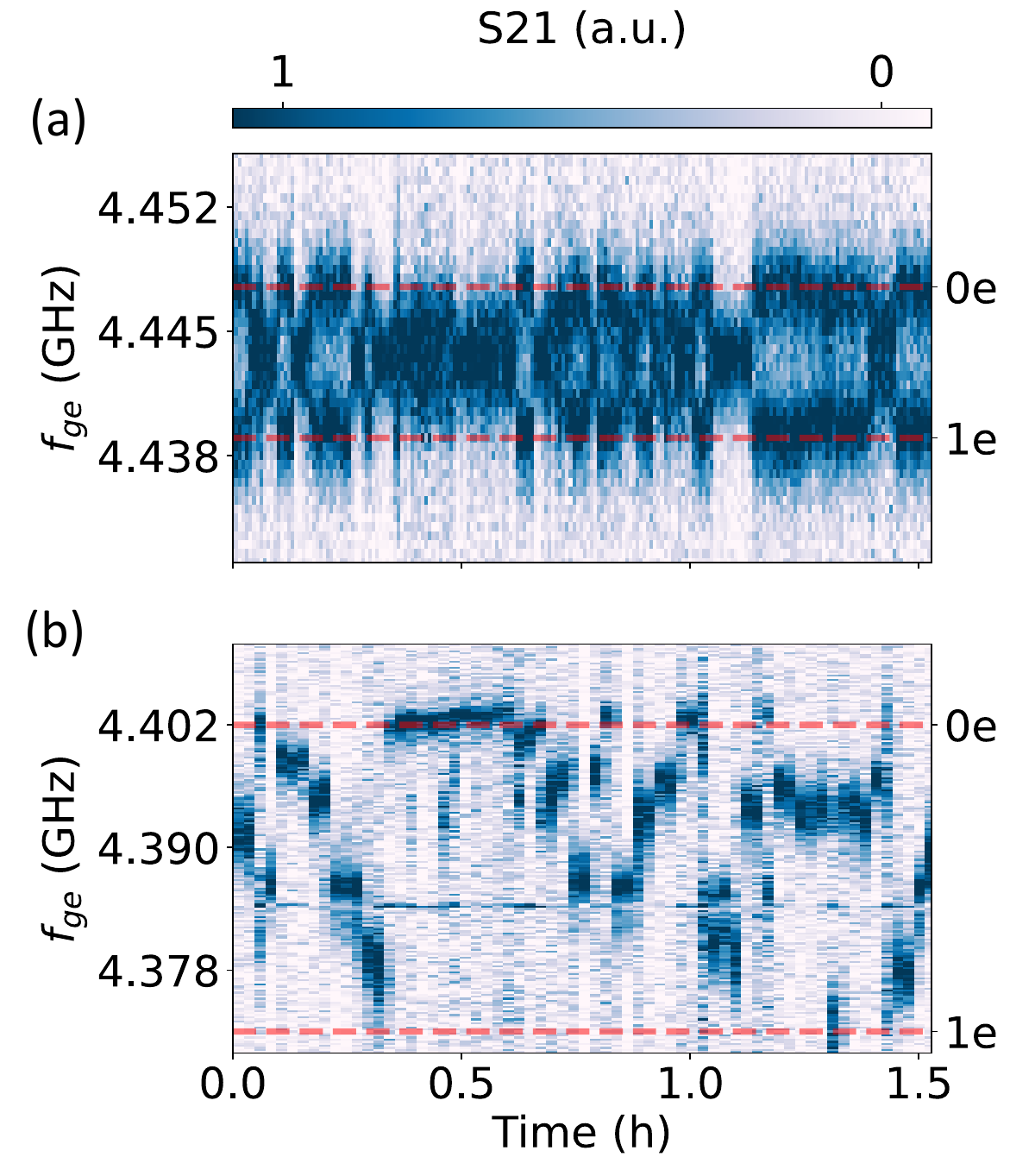}
    \caption{Two-tone spectroscopy of the qubit transition of a non-QP-protected transmon qubit (panel (a)) and a QP-protected transmon qubit (panel (b)), obtained by probing the transmission of the coupled readout cavity. Each pixel is an average of 100 repetitions of qubit measurements ($\sim~200\,\mathrm{ms}$). The horizontal red lines correspond to the $ge$ transition frequencies of the even ($q=0~\mathrm{mod}~2e$) and odd ($q=1~\mathrm{mod}~2e$) charge parity states. The fluctuations of the $ge$ transition frequency are due to random shifts of the offset charges (visible in both panels) and even-odd parity switching (visible only in panel (a)).}
    \label{fig:spec}
\end{figure}

Figure \ref{fig:spec}a shows that for a $\mathcal{NP}$ qubit, two separate qubit resonances are recorded over the duration of each measurement. The splitting between resonances vanishes only at an offset charge $q=0.5e$ (mid-way between the red lines), where the qubit frequency is insensitive to the charge parity switching. Simultaneous observation of two resonances indicates that the charge parity switching occurs at a time scale shorter than the averaging time of a single spectroscopic measurement ($\sim 200\,\mathrm{ms}$). This is consistent with earlier observations of fast tunneling of quasiparticles across the qubit junction (typically, over a time scale much less than 1 ms~\cite{diamond2022} in devices without gap engineering). Random sub-$1e$ shifts of the offset charge at a time scale of a few minutes can be attributed to the coupling of the qubit to fluctuating two-level systems (TLS) in the qubit environment. Similar rates of sub-$1e$ shifts of the offset charge have been observed in prior experiments with transmon qubits \cite{sun2012a}.

The main result of the work is shown in Fig. \ref{fig:spec}b. For the QP-protected qubit ($\mathcal{P}$), every spectroscopic measurement of the resonator records only one qubit resonance, and $f_{ge}$  remains fixed until the next TLS-induced sub-$1e$ shift of the offset charge. We observe that the gap modulation shown in Fig. \ref{fig:design}b suppresses the QP tunneling across a qubit junction over time scales greater than $\sim 10^3\,\mathrm{s}$. The efficiency of gap engineering indicates that most non-equilibrium QPs have energies close to the gap edge. At the same time, the rate of sub-$1e$ shifts of the offset charge, caused by the qubit-TLS coupling, is similar to that in $\mathcal{NP}$ qubits. 

\begin{figure}[t!]
    \centering
    \includegraphics[width=\columnwidth]{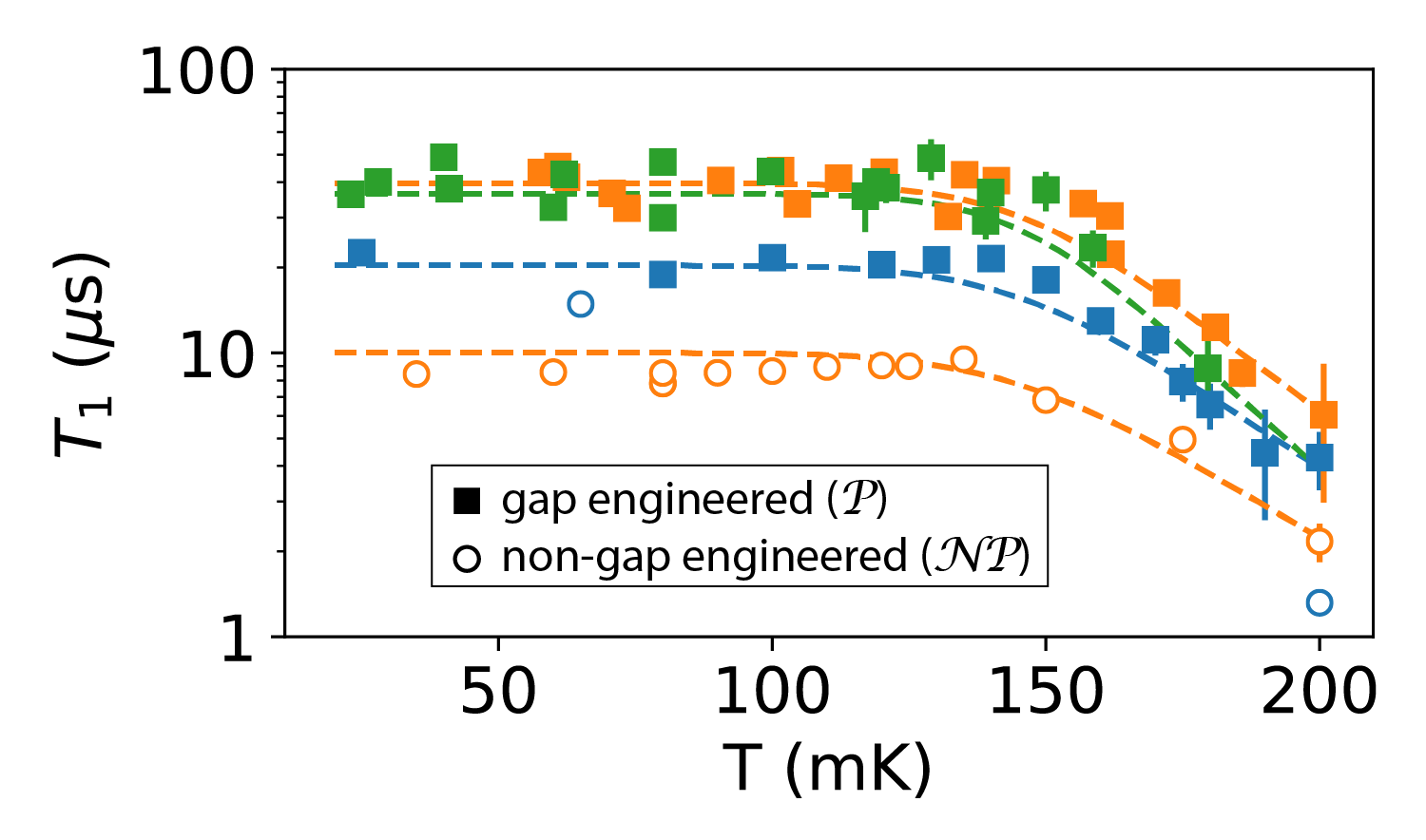}
    \caption{Temperature dependence of the qubit energy relaxation time $T_1$ for devices with gap engineering (represented by squares) and without gap engineering (represented by circles). The device parameters are listed in Table~1. The $T_1$ time is shown versus temperature for QP-protected devices: 1 $\mathcal{P}$ (orange squares), 2 $\mathcal{P}$ (blue squares), and 3 $\mathcal{P}$ (green squares), and non-QP-protected devices: 1 $\mathcal{NP}$ (orange open circles) and 2 $\mathcal{NP}$ (blue open circles). The data is fit to the form expected for inelastic tunneling of thermally-generated QP, as detailed in the text.}
    \label{fig:timedomain}
\end{figure}

We characterized the energy relaxation time $T_1$ for four distinct device categories: charge-sensitive ($E_J/E_C=10-25$) and charge-insensitive devices ($E_J/E_C>50$), for $\mathcal{P}$ and $\mathcal{NP}$ devices. For both charge-sensitive and charge-insensitive qubits, we observed a significant improvement in coherence through gap engineering, which we attribute to the suppression of QP poisoning. For the charge-sensitive devices, time-domain measurements were performed on transitions that corresponded to both even and odd charge parities, with no significant difference in $T_1$.

Figure \ref{fig:timedomain} displays the temperature dependences of $T_1$ for all the devices studied. The lowest measured qubit temperature in QP-protected devices was $49\,\mathrm{mK}$, estimated by using measurements similar to that in \cite{Jin2015-wa, geerlings2013demonstrating}. The energy relaxation time for all the qubits is $T$-independent at $T<120\,\mathrm{mK}$. In this temperature range, we observed a significant (up to a factor of 4) increase in $T_1$ for $\mathcal{P}$ qubits compared to $\mathcal{NP}$ devices. This observation suggests that for the device packaging, filtering, and fridge wiring used in this work, energy relaxation in $\mathcal{NP}$ devices was limited by the inelastic tunneling of NQP. In the QP-protected qubits, QP poisoning is suppressed as indicated by the parity switching measurements. Here, the energy relaxation at low $T$ is expected to be limited by dielectric losses. We observed similar values of $T_1$ for both charge-sensitive and charge-insensitive QP-protected transmon qubits over this temperature range ($\sim40~\mathrm{\mu s}$). Consecutive measurements of $T_1$ over a long time ($\sim$ 1 hour) show that the energy relaxation rate does not fluctuate significantly within the measurement time interval ($\sim 200\,\mathrm{ms}$). 

At higher temperatures, where we observe charge parity violation due to increasing QP density, the $T_1$ decreases rapidly (Fig \ref{fig:timedomain}).  We fit the data assuming that the dominant energy relaxation mechanism in this temperature range is inelastic tunneling of thermally-generated QP and the decay rate $\Gamma_1=1/T_1$ is proportional to the fractional QP density \cite{Glazman2021}:
\begin{equation}
    x_{QP} (T)= x_{NQP} + \sqrt{2\pi T/\Delta } \exp{(-\Delta/T)},
    \label{eq:qptherm}
\end{equation}
The fits shown in Fig.~\ref{fig:timedomain} capture the crossover between this temperature dependence and the thermal-QP independent plateau observed at lower $T$, which serves as a free parameter for the fit of the $T_1 (T)$ dependence. We posit that the transmon decay rate at the lowest temperatures is limited by NQP for the $\mathcal{NP}$ devices, which allows us to infer the non-equilibirium quasiparticle fraction $x_{NQP}$. From the fit, we obtain $T_c = 1.31 \pm 0.04 \mathrm{K}$ and identify a crossover temperature of $169$ mK, where the densities of thermally-excited quasiparticles (QP) and non-equilibrium quasiparticles (NQP) are similar. By applying the crossover temperature to Eq. \ref{eq:qptherm}, we determine the NQP fraction to be $x_{NQP} = 8.0\times 10^{-7}$. This value is consistent with that obtained using the plateau value of the decay rate and Eq. (1) in the Supplement ($x_{\mathrm{NQP}}\sim 1.8\times 10^{-6}$), which assumes the phase fluctuations are small in the calculation of the QP tunneling matrix element. Assuming the density of electron states in Al at the Fermi surface is $\nu_0 = 1.6-1.72 \times 10^{22}/(\mathrm{eV cm^3})$ per spin~\cite{Ptitsina1997-ya,day2003a}, these quasiparticle fractions correspond to an NQP per unit volume of $n_{\mathrm{NQP}} \sim 5-12/\mathrm{\mu m^3}$. While on the high-end, this NQP density is consistent with the range of values ($10^{-6}-10^{-8}$) reported in superconducting qubit experiments~\cite{serniak2018a,day2003a}. (See the Supplementary Material).

Gap engineering cannot protect the junction from tunneling of 
quasiparticles generated within the $3\,\mathrm{\mu m}$-long high-$\Delta$ portion of the bottom electrode. At $T > 150 ~\mathrm{mK}$, the parity violation time becomes shorter than the time of a single measurement, and two qubit resonances are simultaneously observed, similar to that of a $\mathcal{NP}$ qubit. 

The crossover from charge parity conservation at low $T$ to fast parity switching at $T > 150 ~\mathrm{mK}$ has been previously observed for a bifluxon qubit~\cite{kalashnikov2020a}.

As a concluding remark, we place our work in context with related gap engineering studies on transmon qubits~\cite{sun2012b, pan2022engineering}. Ref.~\cite{sun2012b} attempted to suppress QP poisoning in transmons by depositing a thicker Al film \textit{on top} of the JJ electrodes. However, this work reported no noticeable change in parity violation or improvement in $T_1$. We speculate that depositing the thicker film after oxidation introduced an oxide layer on the bottom electrode, thereby disrupting the proximity effect between the Al films. This may have resulted in the bottom electrode being composed of two films with varying critical temperatures, possibly removing the energy barrier crucial for shielding the junction from NQP. Reference~\cite{pan2022engineering} featured junction leads with a larger gap on both sides of the junction, placed atop an optically defined capacitor pad layer with a smaller gap, in a planar geometry. This work demonstrated the mitigation of switching events from photon-assisted tunneling by the addition of a capping chip and gap engineering but still exhibited switching events at intervals of $\sim1~$s.

To summarize, by gap engineering based on the thickness dependence of $\Delta$ in Al films, we were able to realize a potential barrier that prevented tunneling of non-equilibrium quasiparticles across a Josephson junction in the transmon qubit for tens of minutes. Our estimate shows that the QP poisoning rate was reduced by six orders of magnitude, from $\sim10^3\,s^{-1}$ to $\sim10^{-3}\, s^{-1}$. Since the height of the barrier is only $\sim 0.5\,\mathrm{K}$, the dramatic reduction of QP poisoning implies that the energies of majority of NQP are close to the superconducting gap edge, in line with prior experiments~\cite{mcewen2022resolving,wilen2021a}. Suppression of QP poisoning improves the energy relaxation times of both charge-sensitive and charge-insensitive devices.

\section*{Acknowledgements}
The authors acknowledge Jordan Huang for useful discussions and experimental support. This work was supported by the NSF award RAISE-TAQS 1838979 and the ARO award W911NF-17-C-0024. T.D and S.C acknowledge support from the U.S. Department of Energy, Office of Science, National
Quantum Information Science Research Centers, Superconducting Quantum Materials and Systems Center (SQMS) under contract number DE-AC02-07CH11359.

\bibliography{anystyle}

\end{document}


\title{Supplemental Materials}

\subsection{Sample fabrication}

The transmon qubits were patterned using e-beam lithography with a Manhattan pattern design. Prior to the deposition of Al films, the surface of the chip was descummed with a 30W oxygen reactive-ion etch for 20 seconds. Samples were mounted onto a rotatable sample holder fixed at a $45^\circ$ angle with respect to the direction of Al deposition (see Fig.~2 in the main text). The samples were deposited using e-beam evaporation of Al at base pressure below $1\times 10^{-7}$ Torr.

The fabrication steps are depicted in Fig. 2 of the main text. In step 1, we deposited a $40~\mathrm{nm}$-thick underlayer along direction 1. In step 2, after rotating the substrate holder by $45^\circ$, a 25 nm-thick bottom electrode was deposited along direction 2. This film overlaps with the underlayer almost everywhere except for a $3~\mathrm{\mu m}$-long and $0.2~\mathrm{\mu m}$-wide strip that forms the bottom electrode of the JJ. Note that because of shadowing effects, the thickness of the narrow strip could be less ($\sim 20\,\mathrm{nm}$) than that of a wider portion of this film. Due to the thickness dependence of the critical temperature, the thin strip has $T_c \approx 1.6$ K, whereas the rest of the bottom electrode in direct contact with the underlayer has $T_c \sim 1.3$ K \cite{chubov1968a}. It is important to ensure good overlap (within $0.1~\mathrm{\mu m}$, the coherence length in thin Al films) between the underlayer and the wider portion of the bottom electrode, in order to eliminate variations of $\Delta$ within the cross-section of the wider part of the composite bottom electrode.

After step 2, oxygen at a pressure of 0.4 Torr was introduced to the chamber for 20 minutes. This results in a specific resistivity of the junction of  $\sim 900 \mathrm{\Omega \cdot \mu m^2}$. 

In steps 3 and 4, the substrate holder was rotated by $90^\circ$, and fabrication was completed with the deposition of the top electrode (a 60 nm-thick film with $T_c \approx 1.3\,\mathrm{K}$) along two opposite directions 3 and 4 for better step coverage. The Josephson junction was formed between the thin portion of the bottom electrode and the top electrode. The gap profile along the dashed line in Fig.~2b is shown in Fig.~2c of the main text. 

In the QP-non-protected ($\mathcal{NP}$) samples, the underlayer was omitted and the deposition was completed in steps 2-4 described above. In step 2, a 25 nm thick layer of aluminum was deposited in the wider parts of both electrodes and in the narrow part of the bottom electrode.

\subsection{Device simulations}

\begin{figure}
    \centering
    \includegraphics[width=\columnwidth]{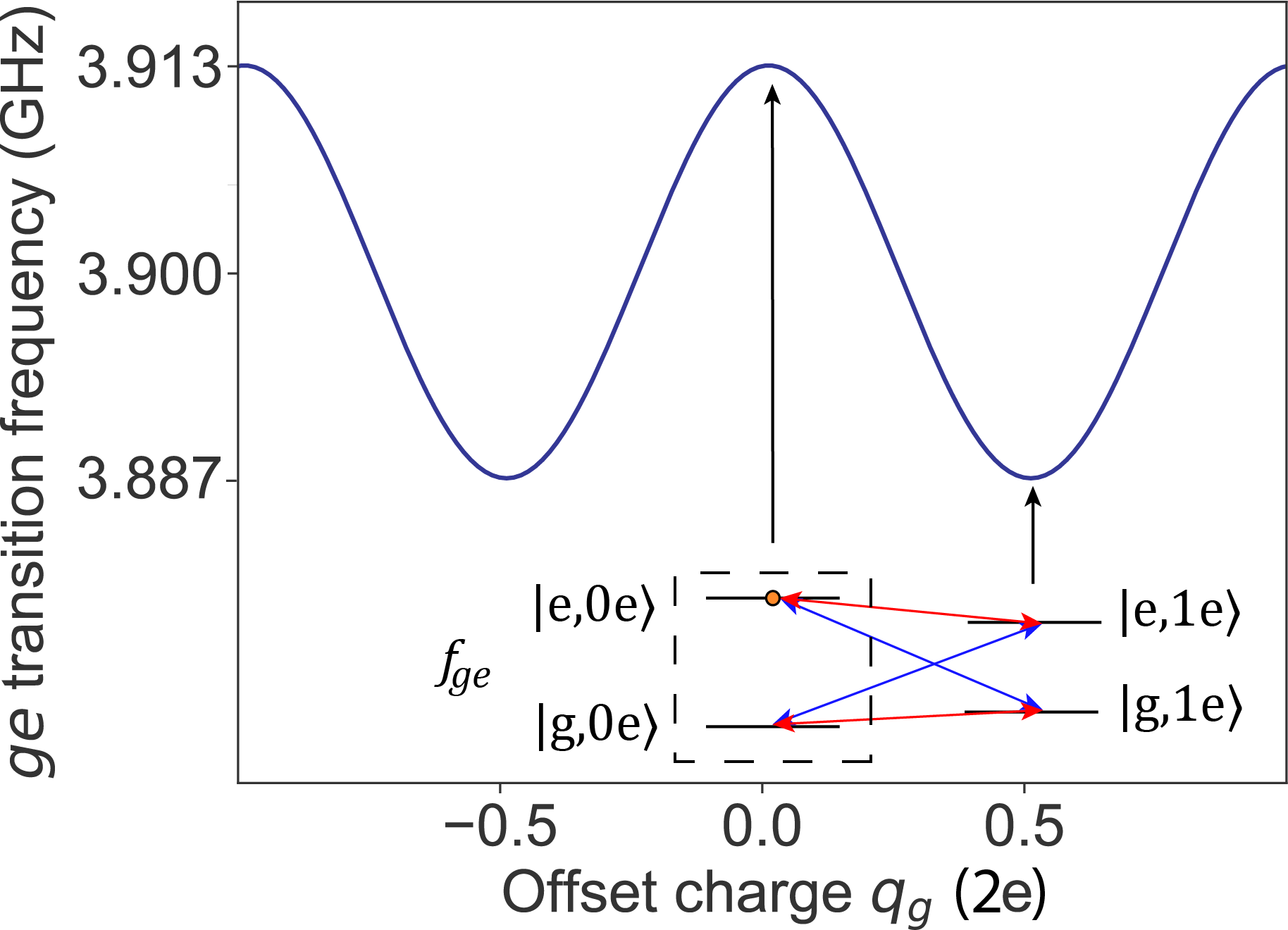}
    \caption{Simulation of the dispersion of the $ge$ transition frequency as a function of offset charge for device $3\mathcal{P}$ in Table 1 of the main text. (Inset) QP jumps (horizontal arrows) lead to changes in the transmon parity ($\ket{0e}$, $\ket{1e}$), changing the $ge$ transition frequency along the dispersion. QP jumps may lead to relaxation or excitation of the qubit excited state (diagonal arrows).}
    \label{fig:charge dispersion}
\end{figure}

We utilized three pad geometries for the transmon qubit, as detailed in Table \ref{table1}, maintaining a pad spacing of $100~\mathrm{\mu m}$ for each configuration. The coupling strength $g$ to the readout cavity was determined using both black-box quantization~\cite{nigg2012black} and the energy-participation method~\cite{minev2021energy}. Subsequently, the transmon-cavity cQED system was modeled using the scqubits package~\cite{groszkowski2021scqubits}. By fitting the measured transmon $ge$ and $ef$ transition frequencies to this model, we extracted the Josephson energy $E_J$ and charging energy $E_C$ for each device; these values are presented in Table \ref{table1}. The $E_J$ was close to that calculated using the Ambegaokar-Baratoff equation and normal-state resistance $R_n$ of the JJ, obtained by room temperature two-probe measurements. The calculated $E_C$ values aligned with those simulated by ANSYS Q3D package to within a $10\%$ margin. The coupling strength $g$ to the readout cavity, as shown in Table \ref{table1}, was determined by comparing the measured and simulated dispersive shifts for the charge-insensitive samples. For charge-sensitive samples, the values were taken directly from simulations, given that the dispersive shift of these devices exhibits a strong dependence on the offset gate charge $q_g$, shown in Supp. Fig 1.

\begin{table*}

\centering
\caption{Device geometry and circuit parameters were measured for all samples. Sample names are denoted by their sample number followed by $\mathcal{P}$ for gap-engineered devices and $\mathcal{NP}$ for non-gap-engineered devices. The spacing between the pads in each device was $100~\mu\text{m}$. The charging energy $E_C$ and the Josephson energy $E_J$ were determined by matching the $ge$ and $ef$ transition frequencies and charge dispersions to simulations conducted using scqubits~\cite{groszkowski2021scqubits}.  The coupling to the readout cavity was simulated using Black box and pyEPR. For charge-insensitive devices, the measured dispersive shift was utilized to extract the $g$. For charge-sensitive samples, both the dispersive shift and the resonator frequency are functions of $n_g$. We denote the expected values of the $\chi$ at $n_g = 0, 0.5$ as $\chi_{0e}, \chi_{1e}$, where the dispersive shift is defined as $2\chi$. $\delta\nu_r = |\nu_r(n_g = 0.5) - \nu_r(n_g = 0)|$ is the dispersion of the readout resonator.} 
\label{device simulation}
\begin{tabular}{|l|ccccccc|}
    \hline  
    Sample 
    & Pad dimensions ($\mathrm{\mu m}$)
    & $E_J (\mathrm{GHz})$ & $E_C (\mathrm{GHz})$ & $g$ ($2\pi \times $MHz) & $\chi_{0e}(2\pi \times\mathrm{MHz})$ & $\chi_{1e}(2\pi \times\mathrm{MHz})$ & $\delta\nu_r$ (kHz)   \\
    \hline \hline
    1 $\mathcal{NP}$ 
    &$600\times 600$
    & 21.67 & 0.150 & 130 & -0.65 (m) &-0.65 (m)& -- \\
    2 $\mathcal{NP}$ 
    &$100\times 600$
    & 7.42 & 0.403  &  122 & -1.63 (s) & -1.23 (s) & 26.7  \\
    \rowcolor{cyan}
    1 $\mathcal{P}$
    &$600\times 600$
    & 13.7 & 0.150 & 153 & -0.55 (m) &-0.55 (m)&  --     \\
    \rowcolor{cyan}
    2 $\mathcal{P}$ 
    &$100\times 600$
    & 6.92 & 0.429  & 122 & -1.93 (s) & - 1.25 (s)& 50.1  \\
    \rowcolor{cyan}
    3 $\mathcal{P}$
    &$100\times 600$ & 5.92 & 0.400 & 122 & -4.05 (s) &-0.96 (s)& 71.2 \\
    \hline
\end{tabular}
\label{table1}
\end{table*}

\subsection{3D cQED setup for transmon measurements}

\begin{figure}
    \centering
    \includegraphics[width=\columnwidth]{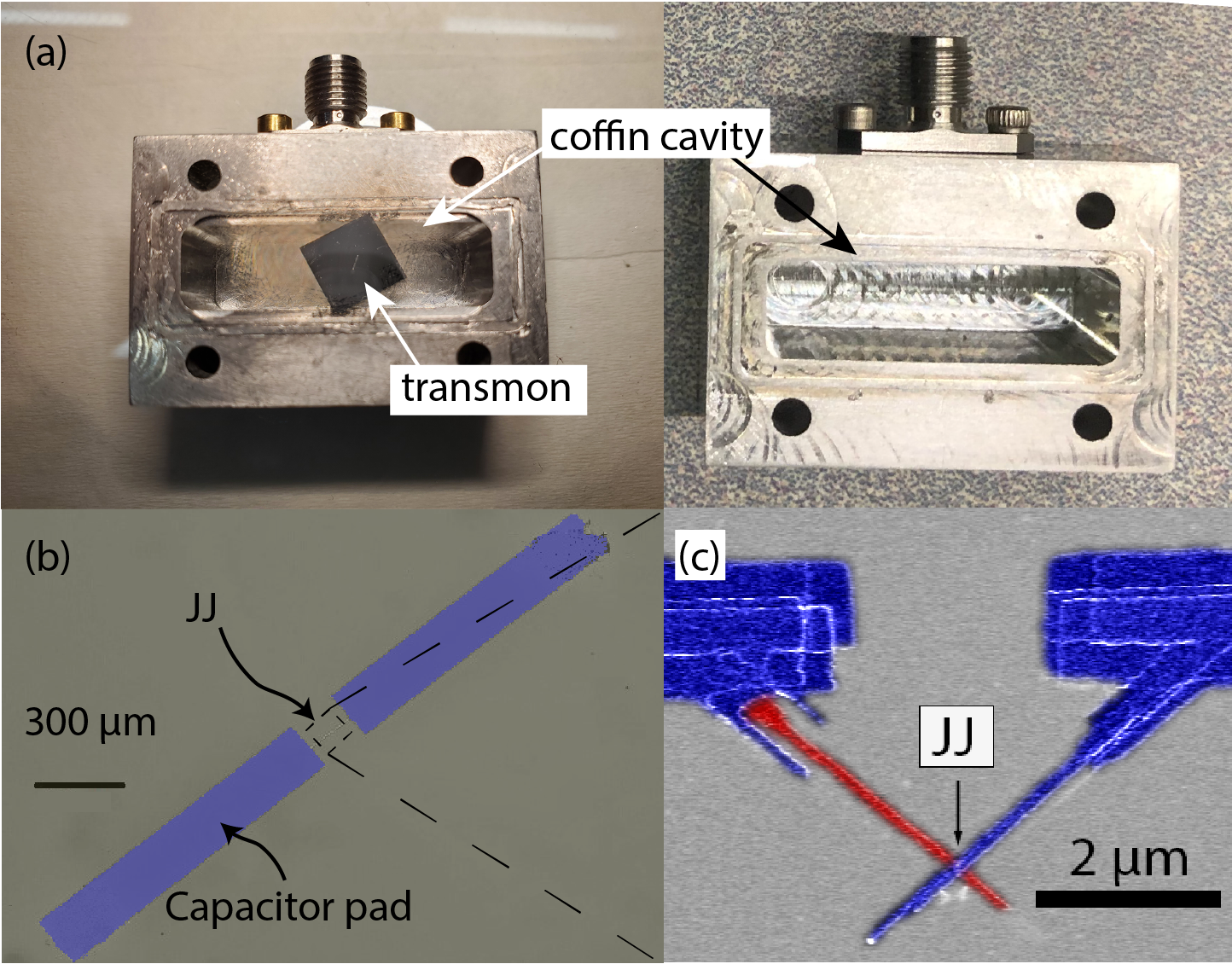}
    \caption{(a) Photograph image of the open coffin style cavity with a 7x7 mm intrinsic silicon chip mounted on the cavity with indium wire. The gap-engineered transmon is in the center of the chip. (b) False color optical microscope image of the 3D transmon with capacitor pads and Josephson junction in the center between the narrow Al leads. The ends of the leads were wire-bonded for two-probe measurements of the JJ normal state resistance. (c) False color micrograph of the Josephson junction shown relative to its position in (b).}
    \label{fig:coffin cavity}
\end{figure}


All studied qubits share the same floating transmon design: a single Al-AlOx-Al Josephson junction flanked by large aluminum capacitor pads was coupled to a three-dimensional (3D) Al readout cavity with a fundamental resonance frequency of $\sim 7.24$ GHz and a loaded quality factor $Q_c\sim 2\times 10^4$. The qubit-resonator coupling ($g$) was chosen to be $120-150$ MHz across the devices, resulting in dispersive shifts as indicated in Table~1 of the Supplementary information. The devices were all characterized in a dilution refrigerator with a base temperature of approximately 25 mK. 

Readout of the transmon was performed by coupling to the fundamental mode of a rectangular waveguide cavity, machined from 6061 Al. The resonator without a transmon chip had a fundamental frequency 7.66 GHz, which reduced to 7.24 GHz with the transmon chip incorporated. The intrinsic
quality factor of the resonator at the base temperature was $Q_i \sim 3 \times 10^5$; the coupling quality factors, controlled by positioning the input and output coupling pins, were $Q_{\mathrm{in}} \sim 2 \times 10^5$ and  $Q_{\mathrm{out}} \sim 2 \times 10^4$, respectively. 

\subsection{Qubit Temperature Measurements}
\begin{figure}
    \centering
    \includegraphics[width=0.5\textwidth]{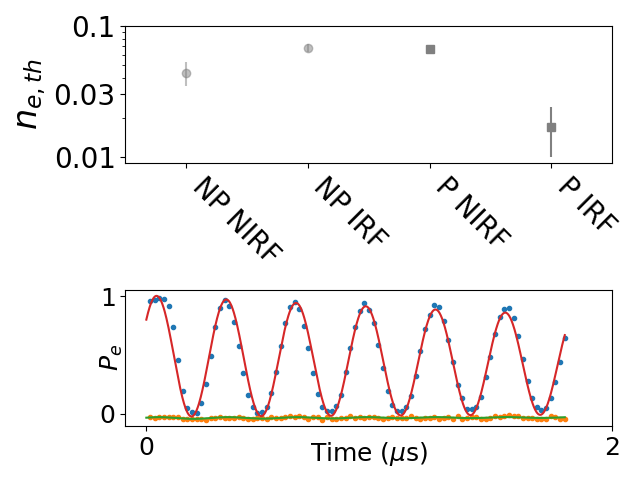}
    \caption{(a) Residual qubit excited state population at a 25 mK base temperature for $\mathcal{NP}$ and $\mathcal{P} $ devices and with (IRF) and without (NIRF) Eccosorb IR filtering of high-frequency radiation. The lowest qubit population was recorded with both QP protection and IR filtering ($\mathcal{P}$ IRF). (b) The qubit population was extracted from a comparison of the $e-f$ Rabi amplitude with (blue points, red fit) and without (orange points, green fit) initializing the qubit with a $g-e$ $\pi$ pulse.}
    \label{fig:Qubit Temp}
\end{figure}

Using the second excited state of the qubit, we obtain a qubit temperature measurement of our device. Quasiparticle tunneling across the JJ may lead to the exciation of the qubit, leading to a residual excited state population~\cite{Jin2015-wa}. The relative amplitude of the $e-f$ Rabi oscillations indicate the population of the excited state, shown in Suppl. Fig. \ref{fig:Qubit Temp}(a). The qubit state population was extracted for similar $\mathcal{NP}$ and $\mathcal{P}$ devices, with high-frequency Eccosorb IR filtering (IRF) and without IR filtering (NIRF). The lowest qubit temperature is measured in devices with both gap engineering and IR filtering. The $e-f$ Rabi oscillations corresponding to this experiment are plotted in Supp. Fig. \ref{fig:Qubit Temp}(b).

\subsection{Calculation of reduced quasiparticle density}

We attribute the improvement in the $T_1$ of the $\mathcal{P}$ devices to the reduction  in the tunneling rate of quasiparticles due to gap engineering. The density of quasiparticles is extracted from the non-gap engineered device $1\mathcal{NP}$. The reduced non-equilibrium quasiparticle density can be extracted from the low temperature value of $T_1$, using Eq. 82 in \cite{Glazman2021}:

\begin{equation}
    \Gamma_{ge,NQP}=32E_J\Big(\frac{\Delta}{2f_{ge}}\Big)^{1/2}\Big(\frac{E_C}{8E_J}\Big)^{1/2}x_{NQP},
\end{equation}

where $\Gamma_{ge,NQP}$ is the contribution to the decay rate of the transmon from NQP tunneling. We obtain a reduced quasiparticle density of $2.6\times 10^{-6}$ using the parameters from Table 1 in the main text.

The reduced NQP density is related to the total NQP density by

\begin{equation}
    x_{NQP}=\frac{n_{NQP}}{2\nu_0\Delta},
\end{equation}

where $\nu_0$ is the density of states of thin film aluminum, $ 1.72 \times 10^{10}\mathrm{\mu m^{-3}eV^{-1}}$ per spin contribution (Ref. 17 in the main text). The density in units of quasiparticles per unit volume obtained from device $1\mathcal{NP}$ is $n_{NQP}=19.6/\mathrm{\mu m^{-3}}$.

\subsection{Fridge wiring and filtering, and thermal characterization}

Measurements were performed in a Bluefors SD-250 dilution refrigerator, which reached a base temperature of 25 mK. Fig.~\ref{wiring diagram} presents the microwave wiring diagram utilized for these measurements. We placed the sample inside a light-tight copper enclosure, which was then situated within a \( \mu \)-metal magnetic shield. Eccosorb filters were added along both the input and output lines just before the SMA ports of the resonator, inside the copper container. These IR filters have an attenuation of \(1-2\,\mathrm{dB}\) at the readout resonator frequency, achieving attenuation $>15$ dB for frequencies above 80 GHz -- twice the superconducting gap. We find these filters assist in decreasing the qubit temperature, likely by reducing photon-assisted quasiparticle creation and tunneling. The lowest qubit temperature was recorded in $\mathcal{P}$ devices with IR filtering, corresponding to an excited state occupation of $1.5\%$ and $T = 49$ mK.

The resonator temperature was found to be primarily limited by thermal photons that enter via the strongly coupled output pin of the cavity. The temperature associated with this port was limited by the thermalization of the last circulator (Low-noise factory) and the attenuator attached to the circulator's third port, which is used to deliver the qubit drive pulses. To minimize the resonator temperature, we used a copper-bodied 20 dB attenuator at this port, which was thermally anchored to the base plate using a copper braid.  

\subsection{Qubit Spectroscopy and Time-domain measurements}

\begin{figure}[h]
    \centering
    \includegraphics[width=\columnwidth]{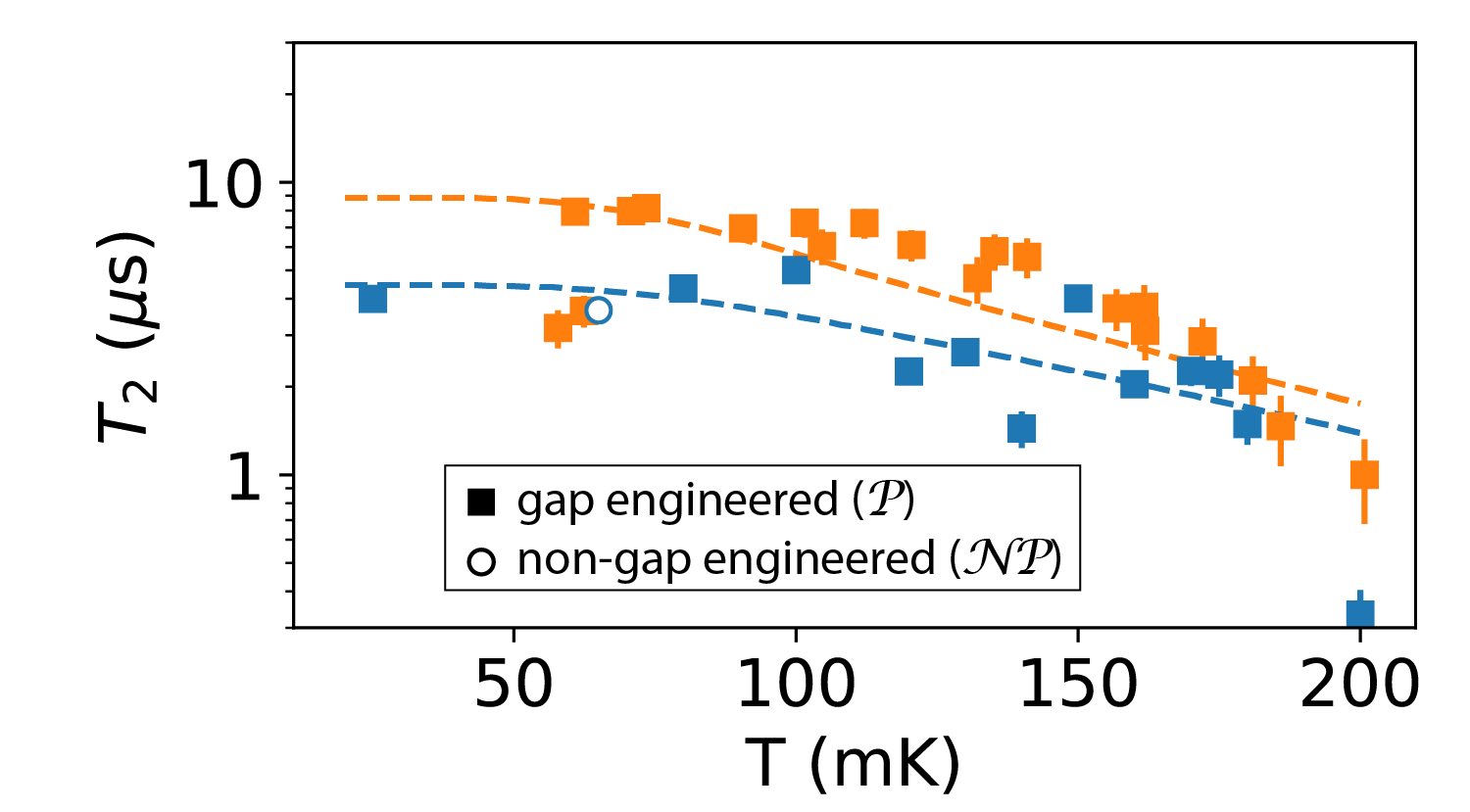}
    \caption{Temperature dependences of the qubit decoherence time $T_2^*$ for QP-protected devices 1$\mathcal{P}$ (orange squares) and 2$\mathcal{P}$ (blue squares), and the non-QP-protected device 1$\mathcal{NP}$ (blue open circles). The dephasing in device 1$\mathcal{P}$ is expected to be limited by thermal photons in the readout resonator, and was fit to the sum of dephasing rates due to energy relaxation and photon shot-noise dephasing. The photon shot noise dephasing is given by SEqn.[1] with an added offset. Device 2$\mathcal{P}$ is additionally limited by charge noise dephasing.
   }
    \label{fig:temp_T2}
\end{figure}

For time-domain data collection, we used a Quantum Machines Operator X (OPX) arbitrary waveform generator. The qubit and readout pulses are produced by mixing the intermediate frequency pulses (bandwidth less than 500 MHz) from the OPX with a 4-7 GHz signal from an Agilent 8257D waveform generator serving as a Local Oscillator (LO). The mixing was performed using a Marki MLIQ-0416 IQ mixer. The resulting pulses are then sent into the fridge. The output from the fridge is amplified and mixed with the reference LO signal to produce I and Q components at an intermediate frequency. These components are then fed into the Quantum Machine OPX ADC and digitally demodulated to obtain the I and Q signals.  Transmon readout was typically performed by probing the readout resonator at the frequency corresponding to the $\ketb{g}$ peak.

The resonance frequency of the qubits was probed as a function of time using two-tone spectroscopy. Each data point in Fig.~3 in the main text  represents the result of averaging 100 individual measurements, with a total acquisition time of $\sim 200$ ms. Fig.~3 shows fluctuations of qubit frequency $f_{ge}$ for two representative transmon qubits: panel (a) - without gap engineering ($\mathcal{NP}$), and panel (b) - with gap engineering ($\mathcal{P}$). The parameters of these transmon qubits are listed in Table 1. Fluctuations of $f_{ge}$ are caused by time variations of the offset charge $q$ due to the QP tunneling ($\delta q=1e$) and slower voltage fluctuations arising from two-level systems (TLS) in the qubit environment ($\delta q<1e$). The horizontal red lines in Figs.~3a,b correspond to qubit frequencies ($f_{ge}$) for even ($q=0~\mathrm{mod}~2e$) and odd ($q=1e~\mathrm{mod}~2e$) charge parities.

For the temperature-dependent measurements conducted in this work, we heated the mixing chamber and allowed the sample to thermalize for 10 minutes. Besides the $T_1$ data presented in the main text, we also recorded the temperature dependence of the dephasing time $T_2^*$ using a standard Ramsey sequence. The results are displayed in Fig.~\ref{fig:temp_T2}. For charge-insensitive devices, the dephasing time is expected to be limited by photon-shot-noise dephasing from thermal photons in the readout cavity, corresponding to a resonator temperature of $86$ mK. Charge-sensitive devices exhibit additional dephasing that we attribute to charge noise. Although we noted a reduction in this dephasing in the gap-engineered device ($2\mathcal{P}$), not all devices show this trend.

The datasets for the temperature-dependence studies were collected independently from the base temperature $T_1$ and $T_2$ data presented in Table 1 of the main text. The fit in Fig.~\ref{fig:temp_T2} for the data associated with sample $1\mathcal{P}$ illustrates the combined dephasing rates due to energy relaxation and photon shot-noise dephasing.  The latter refers to pure dephasing arising from the AC Stark shift induced by thermal photons in the readout resonator and is given by~\cite{clerk2007using}:
\begin{equation}
    \Gamma_\phi (T)=\frac{\kappa}{2}\Re{\sqrt{\Big (1+\frac{2i\chi}{\kappa}\Big)^2+\frac{8i\chi \mathrm{n_{th}}(T)}{\kappa}}-1},
\end{equation}
where $\mathrm{n_{th}} (T)$ is the resonator occupation at temperature $~T$. We use this pure-dephasing rate to extract the resonator temperature in charge-insensitive devices, where we may neglect charge-noise-induced dephasing. The Hahn echo time $T_2^{\text{echo}}$ for device $1\mathcal{P}$ was determined to be $15\,\mu\text{s}$, exceeding the Ramsey $T_2^*$ (6-10 $\mu$s), leading to an additional pure dephasing rate of $\gamma_{\phi} = 56\,\mathrm{kHz}$. With $\chi = 2\pi \times 0.55$ MHz, and $\kappa = 2\pi \nu_r/Q = 2\pi \times 0.36~ $MHz, this leads to a resonator temperature and thermal occupation of $T = 86 $ mK and $n_{\mathrm{th}} = 0.027$, respectively.

\begin{figure*}
    \centering
    \includegraphics[width=0.75\textwidth]{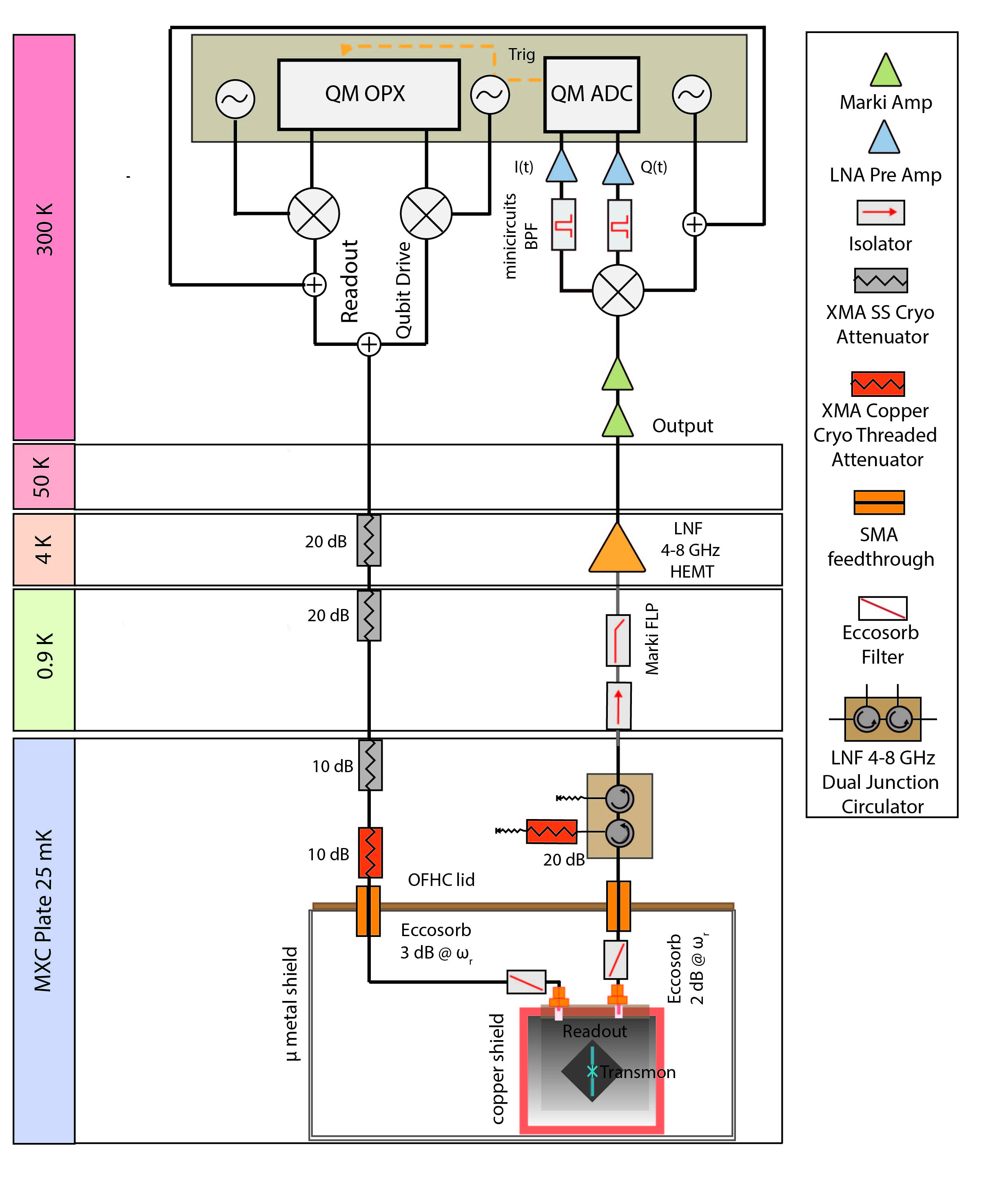}
    \caption{A schematic diagram of the microwave input and output lines in the BlueFors SD250 dilution refrigerator, along with the microwave measurement setup. The Quantum Machines Operator X AWG is used for the RF signal generation (QM OPX) and readout (QM ADC). Eccosorb filters were placed close to the input and output ports of the readout cavity, inside a light tight copper enclosure surrounded by a $\mu-$metal can.}
    \label{wiring diagram}
\end{figure*}

\bibliography{anystyle}